\documentclass[twoside]{article}
\usepackage{fleqn,espcrc2}

\usepackage{epsfig}
\usepackage{fleqn}

\newcommand{\C}{{\sf C\hspace*{-0.9ex}%
    \rule{0.15ex}{1.5ex}\hspace*{0.9ex}}}
\newcommand{\Z}{{\sf Z\hspace*{-0.9ex}%
    \rule{0.15ex}{1.5ex}\hspace*{0.9ex}}}
\def\half{\frac12}
\def\hf{{\textstyle\half}}
\def\nfr#1#2{{\textstyle{\frac{#1}{#2}}}}

\newcommand{\AmS}{{\protect\the\textfont2
  A\kern-.1667em\lower.5ex\hbox{M}\kern-.125emS}}

\title{ADE-quiver theories and Mirror Symmetry\footnote{Talk delivered by U.L.}}
\author{C. Albertsson\address[STHLM]{Department of
Physics, Stockholm University\\ Box 6730,
SE-113 85 Stockholm,
Sweden}\thanks{e-mail: cecilia@physto.se}, B.
Brinne\addressmark[STHLM]\thanks{e-mail: brinne@physto.se}, U.
Lindstr\"om\addressmark[STHLM]
        \thanks{e-mail: ul@physto.se Supported in part by NFR grant 650-1998368 and
by EU contract HPRN-CT-2000-0122.},
        M. Ro\v{c}ek\address{C.N. Yang
Institute for Theoretical Physics,\\ 
                                                                                                                                State University of NewYork,\\ Stony Brook, NY 11794-3840,
USA\\}\thanks{e-mail: rocek@insti.physics.sunysb.edu}, and
       R. von Unge\address{Institute for Theoretical Physics and
Astrophysics\\ Faculty of Science, Masaryk University\\
Kotl\'{a}\v{r}sk\'{a} 2, CZ-611 37, Brno, Czech
Republic\\}\thanks{e-mail: unge@physics.muni.cz Supported by the Czech Ministry of
Education under Contract No 143100006.}
        }

\begin{document}

\begin{abstract}
We show that the Higgs branch of a four-dimensional Yang-Mills
  theory, with gauge and matter content summarised by an $ADE$ quiver
  diagram, is identical to the generalised Coulomb branch of a
  four-dimensional superconformal strongly coupled gauge theory with
  $ADE$ global symmetry. This
  equivalence suggests the existence of a mirror symmetry between the
  quiver theories and the strongly coupled theories.
\vspace{1pc}
\end{abstract}

\maketitle

\section{INTRODUCTION}

The $3D$ mirror symmetry between the Higgs and the Coulomb branch described in
\cite{IS} seems to have a $4D$ counterpart in a mirror symmetry between the Higgs
branch of an $ADE$ quiver gauge theory and the (generalized) Coulomb branch of a
Seiberg-Witten (SW) theory with $ADE$ global symmetry. This symmetry was suggested by
the results of \cite{E6}, where the algebraic curve\footnote{We are not being very
careful with the notation. "Curve" should really be {\em variety}, but we hope that
this will not cause confusion.}  for the
$ADE$ series of four dimensional $ALE$ manifolds was related to the the description
of these manifolds as hyperk\"ahler quotients \cite{LR,HKLR87}. Inclusion of 
Fayet-Iliopolous (FI) parameters in the quotient leads to the deformed
$ADE$-curves, and the curve for $E_6$ surprisingly turned out to be identical to
the SW curve of a  superconformal theory with $E_6$ global symmetry described in
\cite{MN1,MN2,Noguchi}. This agreement between the curves was seen when the FI
parameters were substituted by the Casimirs of the group $E_6$.
The same agreement has since been found for $E_7$ in \cite{E7}
and for $E_8$ in \cite{E8}.

There is thus a strong case for the proposed mirror symmetry and hence for a duality
similar to that in three dimensions. This result is potentially very useful for at
least two reasons. Firstly, the strongly coupled superconformal theories with $E_n$
global symmetry have no Lagrangian description, whereas their mirror images do.
Certain aspects of the theories that are better studied in a Lagrangian formulation
may thus be investigated in the mirror theory. Secondly, since the Coulomb branch
receives quantum corrections but the Higgs branch does not, one consequence of the
mirror symmetry is that quantum effects in one theory arise classically in the dual
theory, and vice versa.

\begin{figure*}
\centerline{\epsfxsize=12cm\epsfbox{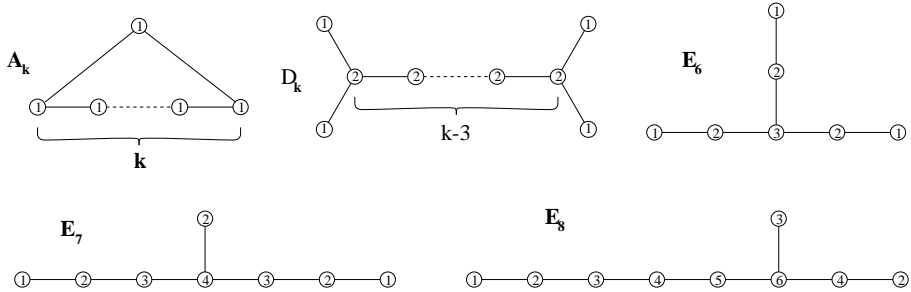}}
\caption{The extended Dynkin diagrams for the $ADE$ quiver theories.}
\label{f1}
\end{figure*}

There is also an advantage in having found the relation between the deformation
parameters of the algebraic curves and the FI parameters. Assume
that one is contemplating a Hanany-Witten (HW) picture of NS5-branes with D4-branes
ending on them as a candidate for the IIA T-dual of
D3-branes on an $E_n$ singularity in IIB. As described in
\cite{E8}, and in section 5 below, moving the NS5-branes in the HW picture
corresponds to blowing up the singularity in the dual picture, the
FI parameters giving the position of the NS5-branes.

  Since we have the relation of
the FI parameters to the parameters governing the deformation of the
algebraic curve, one may now check that the possible motions on the
HW side (allowed by the particular geometry suggested) correspond to
the known allowed deformations on the IIB side. In other words, our results may serve
as a guideline when trying to find a HW picture.

Below we describe the derivation of the results, in particular the ``bug
calculus'' that made the derivation technically feasible. 

\section{The ADE-series}

The quiver theories \cite{DM,JM} are ${\cal{N}}=2$ supersymmetric gauge  theories
that may be characterized by the extended Dynkin diagrams of the $ADE$-series ({\em
quiver diagrams}), as depicted in figure \ref{f1}. Here the gauge
group is
$U(N_1)\times
\ldots
\times U(N_k)$, and the
$i$'th node, labelled by $N_i$ in the Dynkin diagram, corresponds
to a factor $U(N_i)$ in the gauge group. Moreover, each link between two nodes
$i,k$ corresponds to a hypermultiplet in the $(N_{i},\bar{N_{k}})$
representation. The Dynkin diagram
thus sums up both the gauge group and the matter content of the quiver theory. These
theories may be constructed as the worldvolume theory  of $D3$-branes probing an
orbifold singularity. 

There is also a closely related point of view where the Dynkin
diagram represents a hyperk\"ahler quotient by the above gauge group and the
hypermultiplets coordinatize a $4D$ $ALE$-space. It is this latter point of view
which we take as the starting point for our investigations.

The hyperk\"ahler quotient construction \cite{LR,HKLR87} starts from a ${\cal{N}}=2$
supersymmetric nonlinear
$\sigma$ model coupled to an ${\cal{N}}=2$ vector multiplet (including FI terms). In
${\cal{N}}=1$ language the hypermultiplets and vectormultiplets involved are given by
$(z_+,z_-)$ and
$V,S$, respectively, where $z_\pm$ and $S$ are ${\cal{N}}=1$ chiral superfields and
$V$ is an
${\cal{N}}=1$ vector superfield. With the above gauge groups, the quotient found by
integrating out $(V,S)$ produces a new $\sigma$-model with an
$ALE$-space as target space. We shall be particularly interested in the so
called moment map constraints, i.e., the equations that result from integrating
out $S$:
$$
\vbox{\offinterlineskip
\hrule
\halign{&\vrule#&\strut~\hfil#\hfil~\cr
height3pt&\omit&&\omit&&\omit&\cr
&Classi&&Polynomial&&Deformations&\cr
&fication&&&&&\cr
height3pt&\omit&&\omit&&\omit&\cr
\noalign{\hrule}
height3pt&\omit&&\omit&&\omit&\cr
&$A_k$&&$XY-Z^{k+1}$&&$1,\ldots,Z^{k-1}$&\cr
height2pt&\omit&&\omit&&\omit&\cr
&$D_k$&&$X^2+Y^2Z-Z^{k-1}$&&$1,Y,Z,\ldots,Z^{k-2}$&\cr
height2pt&\omit&&\omit&&\omit&\cr
&$E_6$&&$X^2+Y^3-Z^4$&&$1,Y,Z,YZ,$&\cr
&$$&&$$&&$Z^2,YZ^2$&\cr
height2pt&\omit&&\omit&&\omit&\cr
&$E_7$&&$X^2+Y^3+YZ^3$&&$1,Y,Y^2,Z,YZ,$&\cr
&$$&&$$&&$Z^2,Y^2Z$&\cr
height2pt&\omit&&\omit&&\omit&\cr
&$E_8$&&$X^2+Y^3+Z^5$&&$1,Y,Z,YZ,Z^2,$&\cr
&$$&&$$&&$Z^3,YZ^2,YZ^3$&\cr
height2pt&\omit&&\omit&&\omit&\cr}
\hrule}
$$
\vskip -2mm
\centerline{\bf Table 2}
\vskip 3mm
\begin{eqnarray}
z_+ T_A z_-=\, 0~~\qquad &A\notin \hbox{any
$U(1)$ factor}\cr =\, b_A\qquad &A\in\hbox{any $U(1)$ factor},\label{moma}
\end{eqnarray}
where
$b_A$ are FI parameters and $A$ is a group index. Turning off the FI terms results in
the orbifold limit of the $ALE$-space, and conversely non-zero FI terms correspond to
resolutions of the orbifold.
\begin{figure*}
\centerline{\epsfxsize=10cm\epsfbox{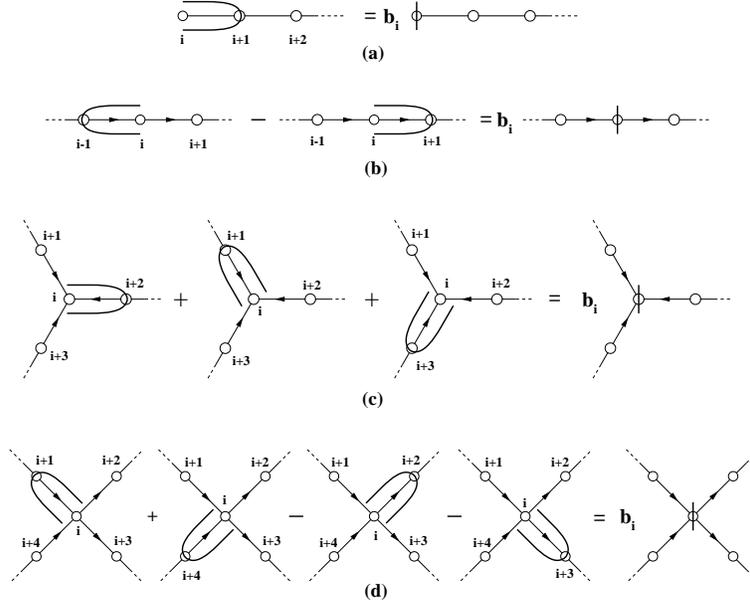}}
\caption{The bug calculus. $b_i$is the fayet-Iliopolous parameter associated with
the 
$i$'th node, and a vertical bar through the $i$'th node represents a $U(N_i)$
Kronecker-$\delta$}
\label{f2}
\end{figure*}

 The $ALE$-spaces, classified by the $ADE$ series, also have a description in terms
of an algebraic curve in $\C^3$ \cite{Aspinwall}. Here the resolution of
the  orbifold
corresponds to certain allowed deformations, as listed in Table 2.

Our goal is to find the relation between the FI parameters
and the deformations of the algebraic curves. The strategy is to form gauge
group invariants from the hypermultiplets and identify those invariants with the
coordinates $X,Y$ and $Z$ of the curves in Table 2. All the calculations should be
done taking the constraints (\ref{moma}) into account. Algebraically finding the
curve with non-zero FI terms is rather a formidable task for most of the models. It
is considerably simplified, however, by use of a ``bug calculus'' \cite{E6}, which we
now describe.

\section{Bug calculus}

\begin{figure}
{\epsfxsize=8cm\epsfbox{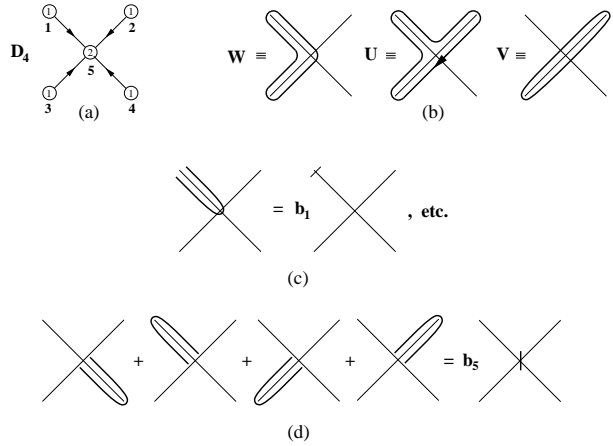}}
\caption{Diagrammatic representation of the $D_4$ invariants and moment map
constraints.}
\label{f3}
\end{figure}

\begin{figure}
{\epsfxsize=8cm\epsfbox{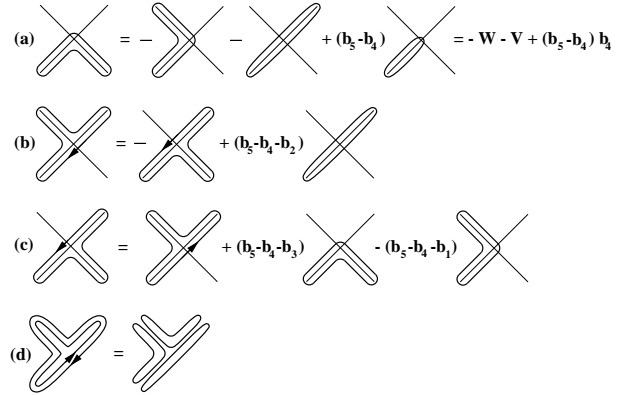}}
\caption{Some typical moves for the $D_4$ example.}
\label{f4}
\end{figure}

We start from the Dynkin diagrams in Table 1 and associate a FI parameter $b_i$ with
the $i$'th node. We also need to keep track of orientation; an arrow from
the fundamental towards the antifundamental representation indicates the way the
chiral field in the hypermultiplet transforms\footnote{A change of direction of an
arrow only affects the final result by a change of sign of the corresponding FI
parameter.}. It is then possible to form matrices from the matter fields and depict
them graphically. 
E.g., the holomorphic constraints (\ref{moma}) can be
represented in bug calculus, each gauge group ({\em{i.e.}}, node) having its own
constraint. For an ``endpoint'' the constraint is shown in figure \ref{f2}a
and for a node in a chain the constraint is shown in figure \ref{f2}b. For nodes
connecting more than two links, the constraints generalize as indicated in figure
\ref{f2}c and d. When manipulating the bugs, the moves are dictated by the moment map
constraints. Using them, one immediately finds that some traces of matrices reduce
to polynomials in
$b_i$ and eventually one is left with a set of non-reducible invariants. Additional
use of the constraints then leads to a relation between (products of) these
variables, which is  the candidate for the curve. For
$D_4$, the Dynkin diagram, the invariants and the constraint are given in figures
\ref{f3}a, \ref{f3}b and \ref{f3}c-d, respectively \cite{E6}. Some of the moves are
described in figure \ref{f4}. Figure \ref{f4}a expresses a four-link diagram in
terms of the
basic four-link
diagrams $W$ and $V$. Figures \ref{f4}b and \ref{f4}c relate $U$ to its
orientation reversed image. Figure \ref{f4}d yields the algebraic curve in
diagramatic form. substituting \ref{f4}a-c and similar relations into \ref{f4}d we
find the curve
\begin{eqnarray}
U^2+U[(b_4-b_1)V+(b_4-b_2)W+a_1]\cr
-W^2V-WV^2+a_2WV=0,
\end{eqnarray}
where $a_1$ and $a_2$ are polynomials in the FI parameters. To find the standard
form of the $D_4$ curve, as given in Table 1, we have to shift the variables
according to
\begin{eqnarray}
U\!\!\!&=&\!\!\!\hf[X+(b_1-b_4)V+(b_2-b_4)W-a_1]\cr
\rule{0mm}{4mm}
V\!\!\!&=&\!\!\!\hf[Y-W+a_2-\hf (b_1-b_4)(b_2-\hf (b_1+b_4))]\cr
\rule{0mm}{4mm} 
W\!\!\!\!&=&\!\!\!\!-Z-\nfr14(b_1-b_4)^2.
\end{eqnarray}
The result is thus an explicit expression for the deformations of the algebraic
curve with (functions of) the  FI parameters as coefficients in the curve.
Modulo the technicalities such as a fair amount  of bug calculus, use
of certain Schouten identities etc, this sums up the procedure for the
$A_k$ and $D_k$ series.

For brevity we do not display the full result here,
 but note that the quantities
that enter in the expression for the deformations in both these series are related
to the  weights of the fundamental representations of the respective Lie algebras,
if we think of each FI parameter as the simple root associated to its node in the
Dynkin diagram. This observation becomes crucial when we turn to the $E$-series,
both for organizing the results in a comprehensive way and for finding the relation
to the SW models.

\section{The  $E_n$ series}

Although more involved, the procedure for deriving the deformations of the
$E_n$-curves follows the lines described in section 3 \cite{E6}-\cite{E8}. The key
to understanding the initially not very illuminating results is to first find an
expression for the Casimirs of the Lie algebras in terms of the FI parameters, and
then invert this. The final expressions for the deformed curves in terms of these
Casimirs are manageable and in fact known; they are the SW curves for the
superconformal ``fixed point'' theories described in \cite{MN1,MN2,Noguchi}. 

The $i$'th Casimir $P_i$ can be found as the coefficient of $x^{d_n-i}$ in the
polynomial
\begin{equation}
det(x-v\cdot H),
\end{equation}
where $d_n$ is the dimension of the fundamental representation of $E_n$ and $v$ is
an arbitrary vector in the Cartan subspace. 
The matrix $v\cdot H$ is given in terms of the weights $\lambda$ of this
representation as $v\cdot H=diag(v\cdot \lambda_1...v\cdot \lambda_d )$. Using that
each FI parameter $b_i$ can be thought of as the scalar product between $v$ and its
corresponding root we rewrite the weights in terms of the FI parameters. This yields
the relation between the $P_i$'s and the $b_i$'s we were looking for.

When comparing our results to the SW curves, the most immediate comparison is with
\cite{Noguchi} where the curves are given in terms of the Casimirs. On the other
hand, the expressions in \cite{MN1,MN2} are in terms of mass-parameters $m_i$ and the
relation
$P_i=P_i(m)$ thus gives us an interpretation of the $b_i$'s in terms of mass
parameters.

While the above description of the derivation gives the principles of the
procedure, there are many techincal obstacles, most notably in the $E_8$
calculation \cite{E8}. In fact, although the bug calculus is very efficient (many
pages of algebra are replaced by a few figures) it was not by itself enough to  
allow us to
perform the $E_8$ calcualtion. Firstly we had to perform the calculations using a
computer program (MAPLE), and secondly we could not do all of the
comparison to the SW curve explicitly. To deal with some of the highest order terms
(e.g., a polynomial of order 30 in eight variables) we had to resort to numerical
methods: inserting random prime numbers we found that also the terms most difficult
to compare agreed. It is thus clear that the $ADE$ quiver theories away from the
orbifold limit have deformed algebraic curves that are identical to the SW
curves of certain superconformal theories with the corresponding global symmetries.

The interpretation of this fact, suggested in \cite{E6}, substantiated in \cite{E7}
and further discussed in \cite{E8}, is the existence of a mirror symmetry between
the Higgs branch of the quiver theories and the Coulomb branch of the SW theories
similar to that which exists for $3D$ gauge theories \cite{IS}. A problem here,
though, is that the Higgs branch is a hyperk\"ahler manifold, whereas the Coulomb
branch in $4D$ is not (in general). This is overcome by the proposal in \cite{E7}
that what is involved  is the {\em generalized} Coulomb branch, defined to be the
$4D$ elliptically fibered space obtained by fibering the SW torus over the usual
Coulomb branch.

\section{Geometrical interpretation of $b_i$}

We close this presentation with a section quoted from \cite{E8} on the geometrical
meaning of the FI terms.

The $A_{n-1}$ quiver theories can be viewed as the world-volume theory of $D3$-branes
on a $\C^2/\Z_n$ orbifold singularity in type IIB string theory. This picture is
T-dual to a picture of type IIA string theory in a background of D4-branes
stretching between NS5-branes
\cite{KLS}. This dual picture, the Hanany-Witten (HW) picture
\cite{HW}, provides an intuitive geometric interpretation of blow-ups
of $A_{n-1}$ type singularities. An analogous picture exists for $D_n$
type singularities \cite{Kapustin,HZ}, and it seems plausible that
there are generalizations also to $E_6$, $E_7$ and $E_8$.  In this
section, we analyze the HW picture for the $\C^2/\Z_n$ case along the
lines of \cite{KLS} (see also \cite{Dasgupta2,Johnson}); in particular
we clarify the role of the Fayet-Iliopoulos terms.

Starting from the type IIB string theory configuration ($\times$ means
the object is extended in that direction, and $-$ means it is
point-like)
\begin{eqnarray*}
 \begin{array}{|c|c|c|c|c|c|c|c|c|c|c|}
         \hline
         & $0$ & $1$ & $2$ & $3$ & $4$ & $5$ & $6$ & $7$ & $8$ & $9$ \\\hline
    \mathrm{sing} & \times & \times & \times & \times & \times & \times & - & - & - & - \\\hline
    \mathrm{D}3   & \times & \times & \times & \times & - & - & - & - & - & - \\\hline
  \end{array}
\end{eqnarray*}
we T-dualize along the 6-direction to get
\begin{eqnarray*}
 \begin{array}{|c|c|c|c|c|c|c|c|c|c|c|}
         \hline
         & $0$ & $1$ & $2$ & $3$ & $4$ & $5$ & $6$ & $7$ & $8$ & $9$ \\\hline
    \mathrm{NS}5  & \times & \times & \times & \times & \times & \times & - & - & - & - \\\hline
    \mathrm{D}4   & \times & \times & \times & \times & - & - & \times & - & - & - \\\hline
  \end{array}
\end{eqnarray*}
in type IIA string theory in flat space-time. There are $n$ NS5-branes,
which all coincide in the 789 directions, but not necessarily in the
6-direction.  Between them D4-branes are suspended, which are the
T-duals of the IIB D3-branes.  The rotational symmetry $SO(3) \simeq
SU(2)$ of the 789 coordinates translates into the $SU(2)_R$ symmetry
of the gauge theory living on the D4-branes.  The hypermultiplets
arise from fundamental strings stretching across the NS5-branes,
between neighboring D4-branes.

Resolving singularities in the IIB picture corresponds to separating
NS5-branes along the 789 directions in the IIA picture. By an $SU(2)$
rotation we can always pick the direction of displacement to be $x^7$.
Note that such a displacement breaks the 789 rotational symmetry; that
is, blowing up a singularity breaks the $SU(2)_R$ symmetry.
If we move some of the NS5-branes in this way, with
the D4-branes still stuck to them, and then T-dualize along $x^6$
again, we do not regain the D3-brane picture. Rather, the now tilted
D4-branes dualize to a set of D5-branes (with nonzero B-field) with
their 67 world-volume coordinates wrapped on 2-cycles. Shrinking these
2-cycles to zero size, each of the wrapped D5-branes is a fractional
D3-brane, which cannot move away from the singularity.  Thus a
fractional D3-brane corresponds to a D4-brane whose ends are stuck on
NS5-branes.

To move a fractional D3-brane, or, equivalently, a wrapped D5-brane,
along the 6789 directions, we need to add $n-1$ images (under $\Z_n$),
all associated with a 2-cycle each. The sum of the full set of
2-cycles is homologically trivial and can be shrunk to zero size. Then
the collection of wrapped D5-branes will look like a single D3-brane
that can move around freely in the orbifold.  This procedure
corresponds in the HW picture to starting out with a single D4-brane
stretching between two of the $n$ NS5-branes, and wanting to move the
D4-brane (in the 7-direction, say) away from the NS5-branes, detaching
its ends. In order not to violate the boundary conditions of the
D4-brane, we then need to put one D4-brane between each unconnected
pair of NS5-branes and join them at the ends. We then get a total of
$n$ D4-branes forming a single brane winding once around the periodic
6-direction.  The D4-brane may now be lifted off the NS5-branes and
can move freely, corresponding to the free D3-brane in the T-dual picture.

We may also gain some insight concerning the role played by the FI
parameters in the HW picture, from the world-volume theory of a wrapped
D5-brane on the orbifold singularity. Consider such a brane living in
the 012367 directions, with its 67 world-volume coordinates wrapped on
a 2-cycle $\Omega_k$.  The Born-Infeld and Chern-Simons terms in the
world-volume action are, schematically, \cite{DM}
\begin{eqnarray}
  \nonumber I_{D5} &=& \int d^6x \sqrt{\det(g+{\cal F})}
  +\mu\int C^{(6)}\\
 && +\nonumber \mu\int C^{(4)}\wedge {\cal F} 
   +\mu\int C^{(2)}\wedge {\cal F}\wedge {\cal F}\\
 && +\mu\int C^{(0)}\wedge {\cal F}\wedge {\cal F}\wedge {\cal F} ,
  \label{eq:D5action}
\end{eqnarray}
where $g$ is the metric on the world-volume, $C^{(p)}$ is the R-R
$p$-form, $\mu$ is a constant, and ${\cal F}=F^{(2)}+B^{NS}$ where the
2-form $F^{(2)}$ is the field strength of the gauge field on the brane
and $B^{NS}$ is the NS-NS 2-form on the brane.  Dimensional reduction
to the 0123 directions, by integrating over the 2-cycle, puts the
first term of (\ref{eq:D5action}) on the form
\begin{equation}
  \int_{\Omega_k} d^2x \sqrt{\det( g_2+ {\cal F}_2)}
  \int d^4x \sqrt{\det( g_4+ {\cal F}_4)} ,
  \label{eq:D5BIred}
\end{equation}
where ${\cal F}_2 = C^{(2)}+B^{NS}$, $g_2$ is the metric on the 67
directions, and $g_4$ is the metric on the 0123 directions.  Expanding
(\ref{eq:D5BIred}) we obtain the coupling constant $g_k^{-2}$ in four
dimensions as the coefficient of $\int d^4x F_{\mu\nu}F^{\mu\nu}$.  It
is just the factor on the left in (\ref{eq:D5BIred}), which we can
write as
\begin{equation}
  g_k^{-2} = \left| \int_{\Omega_k} \left( B^{NS} + iJ \right) \right| .
  \label{eq:gk}
\end{equation}
In the HW picture the coupling constant of the four dimensional theory
is proportional to the length of the D4-brane in the 6-direction. Hence
(\ref{eq:gk}) measures the total distance between two NS5-branes between which the
D4-brane is suspended. Furthermore, since the distance between the NS5-branes in
the isometry direction (in our case $x^{6}$) is given by the flux of
the $B^{NS}$ field on the corresponding cycle, we have to interpret
$\int_{\Omega_k} J$ as the position of the NS5-branes in a direction
orthogonal to that, let us choose $x^{7}$. Movement of the NS5-branes
in the remaining directions $x^{8}$ and $x^{9}$ now corresponds to
turning on the $SU(2)_R$ partners of the K\"{a}hler form.

The integral of $J$ over a 2-cycle is also, by definition, a
Fayet-Iliopoulos term.  A hyperk\"ahler manifold has an SU(2) manifold
of possible complex structures.  Choosing a complex structure we can
define the K\"{a}hler form $J$ as $\omega^1$, and the holomorphic
2-form as $\omega^2 +i\omega^3$. These three 2-forms rotate into each
other under $SU(2)_R$ transformations, corresponding to choosing a
different complex structure. The $k$:th triplet of FI terms is defined
by the period of $\vec{\omega} = (\omega^1,\omega^2,\omega^3)$ (and
hence also transforms as a triplet under $SU(2)_R$), as
$$
\vec{\zeta}_k  \equiv \int_{\Omega_k} \vec{\omega} .
$$
Hence
$$
\zeta^R_k = \int_{\Omega_k} J ,
$$
where $\zeta^R_k$ is the real component of the triplet of FI terms
$\vec{\zeta}_k = (\zeta_k^R,\zeta_k^C,\overline{\zeta_k^C})$.

Another way to obtain the FI terms of the four-dimensional Yang-Mills
theory is via dimensional reduction and supersymmetrization of the
D5-brane world-volume theory \cite{DM}.  The third term of
(\ref{eq:D5action}) can be rewritten as
$$
\int d^6x (A_\mu - \partial_\mu c^{(0)})^2,
$$
where $c^{(0)}$ is the Hodge dual potential of $C^{(4)}$ in six
dimensions.  After integration over the $k$:th 2-cycle we
supersymmetrize this to
$$
\int d^4x d^4\theta ({\mathbf C}_k - \overline{\mathbf C}_k -
{\mathbf V})^2,
$$
where ${\mathbf C}_k$ is a chiral superfield whose complex scalar
component is $c^{(0)} + i\zeta^R_k$, and $\mathbf V$ is the vector
superfield containing $A_\mu$.  Here the imaginary part $\zeta^R_k$ of
the scalar component is the real FI term in four dimensions, and
we see that it arises as the superpartner of $c^{(0)}$.

\section{Conclusions}

As mentioned in the introduction, the mirror symmetry found is useful because
it relates quantum and classical regimes as well as theories without a Lagrangian
formulation to theories with such a formulation. Also mentioned is that the
geometrical interpretation of the FI terms may serve as a guide-line for finding
dual HW pictures of the $D3$-branes on the $E_n$ singularities.

In \cite{E6} it is shown that higher dimensional hyperk\"ahler
quotients may also be related to quiver diagrams, although the connection to the
simple Lie algebra classification is lost. In particular, several four (complex)
dimensional spaces were constructed. This opens up the possibility of a
systematic investigation of these spaces, perhaps leading to an eventual
classification. The physical relevance of such spaces is not obvious, but perhaps
they have a place in an $F$-theory picture.


\begin{thebibliography}{77}

\bibitem{E6}U.~Lindstr\"om, M.~Ro\v{c}ek and R.~von Unge,
\emph{Hyperk\"ahler Quotients and Algebraic Curves}, JHEP {\bf 0001}
(2000) 022, hep-th/9908082
\bibitem{E7}I.Y.~Park and R.~von Unge,
\emph{Hyperk\"ahler Quotients, Mirror Symmetry, and F-theory}, JHEP {\bf 0003}
(2000) 037, hep-th/0001051
\bibitem{E8}
C.~Albertsson, B~Brinne, U.~Lindstr\"om, and R.~von Unge,
``E8 Quiver Gauge Theory and Mirror Symmetry.,''
[hep-th/0102038].
\bibitem{IS}K.~Intriligator and N.~Seiberg,
\emph{Mirror Symmetry in Three Dimensional Gauge Theories}, Phys.~Lett.~{\bf B387}
(1996) 513-519, hep-th/9607207
\bibitem{HW}A.~Hanany and E.~Witten,
\emph{Type IIB superstrings, BPS monopoles, and three-dimensional gauge dynamics}, Nucl.~Phys.~{\bf B492}
(1997) 152-190, hep-th/9611230
\bibitem{JM}C.V.~Johnson and R.C.~Myers,
\emph{Aspects of Type IIB Theory on ALE Spaces}, Phys.~Rev.~{\bf D55}
(1997) 6382-6393, hep-th/9610140
\bibitem{DM}M.R.~Douglas and G.~Moore,
\emph{D-branes, Quivers, and ALE Instantons}, hep-th/9603167
\bibitem{LR}U.~Lindstr\"om and M.~Ro\v{c}ek,
\emph{Scalar Tensor Duality and N=1,2 Nonlinear $\sigma$-models}, Nucl.~Phys.~{\bf B222}
(1983) 285


\bibitem{HKLR87}N.J.~Hitchin, A.~Karlhede, U.~Lindstr\"om and
M.~Ro\v{c}ek, \emph{Hyperk\"ahler Metrics and Supersymmetry}, Commun.~Math.~Phys.~{\bf 108}
(1987) 535-589
\bibitem{Aspinwall}P.S.~Aspinwall, \emph{K3 Surfaces and String Duality},
Lectures given at the TASI-96
summer school on Strings, Fields and Duality,
hep-th/9611137
\bibitem{MN1}
J.~A.~Minahan and D.~Nemeschansky,
``An N = 2 superconformal fixed point with E(6) global symmetry,''
Nucl.\ Phys.\ B {\bf 482}, 142 (1996)
[hep-th/9608047].
\bibitem{MN2}J.~Minahan and D.~Nemeschansky,
\emph{Superconformal Fixed Points with E$_n$ Global Symmetry}, Nucl.~Phys.~{\bf B489}
(1997) 24-46, hep-th/9610076
\bibitem{Noguchi}M.~Noguchi, S.~Terashima and S-K.~Yang,
\emph{N=2 Superconformal Field Theory with ADE Global Symmetry on a D3-brane Probe}, Nucl.~Phys.~{\bf B556}
(1999) 115-151, hep-th/9903215
\bibitem{KLS}A.~Karch, D.~L\"ust and D.J.~Smith,
\emph{Equivalence of Geometric Engineering and Hanany-Witten via Fractional Branes}, Nucl.~Phys.~{\bf B533}
(1998) 348-372, hep-th/9803232
\bibitem{Kapustin}A.~Kapustin,
\emph{D$_n$ Quivers From Branes}, JHEP {\bf 9812}
(1998) 015, hep-th/9806238
\bibitem{HZ}A.~Hanany and A.~Zaffaroni,
\emph{Issues on Orientifolds: On the brane construction of gauge theories with $SO(2n)$ global symmetry}, JHEP {\bf 9907}
(1999) 009, hep-th/9903242
\bibitem{Dasgupta2}K.~Dasgupta and S.~Mukhi,
\emph{Brane Constructions, Fractional Branes and Anti-deSitter Domain Walls}, JHEP {\bf 9907}
(1999) 008, hep-th/9904131
\bibitem{Johnson}C.V.~Johnson,
\emph{D-Brane Primer}, 
Lectures given at ICTP, TASI, and BUSSTEPP, Sec.~9, hep-th/0007170


\end{thebibliography}
\end{document}